\def\order{{\it O}}

\def\qbar{\overline{q}}

\documentclass{elsart}

\begin{document}
\begin{frontmatter}
\title{Glue in the light-front pion}
\author{Robert J. Perry}
\address{Department of Physics, The Ohio State University,
Columbus, OH 43210 USA, perry.6@osu.edu}
\date{25 August 2000}


\begin{abstract}

It may be possible to approximate the full pion wave function in light-front
QCD using only $q \qbar$ and $q \qbar g$ Fock space components. Removing zero
modes and using invariant-mass cutoffs that make a constituent approximation
possible leads to non-canonical terms in the QCD hamiltonian, and forces us
to work in the broken symmetry phase of QCD in which chiral symmetry breaking
operators must appear directly in the hamiltonian because the vacuum is
trivial. Assuming any candidate interactions are local in the transverse
direction, I argue that they are probably relevant operators so that
perturbation theory is not modified at high energies. Since light-front
chiral symmetry corresponds to quark helicity conservation, we can readily
identify candidate chiral symmetry breaking interactions. The only candidate
relevant operator is the quark-gluon emission/absorption operator with a quark
spin flip. I argue that this operator can only produce the physical
$\pi-\rho$ mass splitting if the $q \qbar g$ component of the pion is
significant.

\end{abstract}
\end{frontmatter}


\section{A Constituent Pion?}

Can the pion be adequately approximated as a quark-antiquark bound state? The
constituent quark model (CQM) works well for massive hadrons, but most
theorists believe that the pion lies outside the CQM's range of ``validity."
Early attempts to relate the CQM to QCD focussed on the relationship between
current and constituent quarks, advocating a picture in which the constituent
quarks are dressed current quarks; but only naive transformations were
investigated and gluons played little or no role in these investigations.

I believe that a constituent approximation may arise in light-front QCD, as
was most convincingly argued by Brodsky, Lepage and
collaborators \cite{lfreview}. However, I believe the pion will stand out
because it contains a significant amount of glue. Lepage, Brodsky, Huang, and
Mackenzie showed that under reasonable assumptions the probability of finding
a $q \qbar$ component in the pion is of order $1/4$; however, their arguments
can be turned around to show that higher Fock space components are important.
Moreover, their estimate depends on a model wave function that is not derived
from QCD and it is possible that this probability is even higher than $1/4$.
It should not be surprising that components with glue are large in the pion.
What is amazing is that the $q \qbar$ component is sizable.

I cannot review the computational scheme that
might be used to produce a massless constituent pion \cite{perry1}. Several
theorists, guided by Ken Wilson, have developed renormalization techniques
that can be used to attack QCD bound state problems in much the same way as
QED bound states can be treated in Coulomb gauge. They first argued that
constituent masses and confinement should be added by hand for all partons,
and that the strong-coupling, relativistic limit of such models might be full
QCD \cite{lfqcdwilson}. However, I have shown that confinement does not have
to be added by hand, because it appears in the renormalized light-front QCD
hamiltonian at second order \cite{perry1}. Moreover, I have argued that this
confining interaction can produce the constituent mass scale as happens in the
MIT bag model. Martina Brisudova has shown that the second order hamiltonian
produces reasonable results for heavy mesons \cite{brisudova} and Brent Allen
has shown that it also produces reasonable glueball results \cite{allen}, {\it
without adding any non-canonical parameters to QCD}.

To obtain a constituent approximation, we remove zero modes and are forced to
work in the broken symmetry phase of the theory. Interactions that explicitly
violate chiral symmetry should be induced by spontaneous chiral symmetry
breaking, and in general the new symmetry breaking couplings must be adjusted
by hand to obtain the correct $\pi - \rho$ mass splitting. 

There are many non-canonical operators induced by renormalization (e.g., a
gluon mass term), and in principle the dimensionless couplings for the
relevant, $\mu$, and marginal, $\gamma$, operators act as independent
parameters that must be tuned to fit data or recover symmetries. We assume
that all couplings induced by renormalization are functions of the canonical
QCD coupling and that they vanish when this coupling goes to zero. In other
words, we allow new relevant couplings, $\mu(g_\Lambda)$ with $\mu(0)=0$, and
marginal couplings, $\gamma(g_\Lambda)$ with $\gamma(0)=0$. These conditions
constitute coupling coherence (or coupling reduction), and we have shown that
they produce effective interactions that restore all broken symmetries at
leading orders in perturbation theory. I note that for an asymptotically free
theory the boundary conditions imply that $\mu_\Lambda
\rightarrow 0$ and $\gamma_\Lambda \rightarrow 0$ as $\Lambda \rightarrow
\infty$. 

I assume that the
new chiral symmetry breaking couplings are fixed functions of the canonical
coupling, but there is no reason to expect perturbative renormalization group
equations to produce the correct dependence, because they are probably
non-analytic functions of the coupling.\footnote{It is interesting to note
that a perturbative RG analysis does lead to a relevant chiral symmetry
breaking coupling that scales like $exp\{-1 / (\beta g^2)\}$, but only in
asymptotically free theories.}

A marginal operator introduced at low energies will tend to have a comparable
effect when the cutoff is increased, and it will typically affect
perturbation theory at some finite order. Only relevant operators can be
introduced at low cutoffs and necessarily have no effect at any finite order
of perturbation theory at higher cutoffs. The only candidate relevant
operator is gluon emission/absorption with a quark spin flip, shown
below.

It seems that the only simple way in which a constituent
picture that includes confinement and chiral symmetry breaking effects is for
this single relevant interaction to produce the entire $\pi-\rho$ mass
splitting. I assume that the vertex mass, which is $g m_F$ to
leading order in perturbation theory, becomes $M_{\chi SB}$ because of
spontaneous symmetry breaking.

\section{Chiral symmetry in light-front field theory}

I first review chiral symmetry without zero modes in light-front QCD,
following Mustaki \cite{mustaki}. The fermion field naturally separates into
two-component fields
$\psi = \psi_+ + \psi_-$ on the light-front, where $\psi_+$ is dynamical and
$\psi_-$ is constrained.  The light-front chiral transformation applies
freely only to the two-component field $\psi_{\pm} = \frac{1}{2}
\gamma^0 \gamma^\pm \psi$, because the constraint
equations are inconsistent with the normal chiral transformation rules.  The
minus-component $\psi_-$ is determined by the light-front constraint
\begin{equation}
	\psi_- = { 1 \over i \partial^+} \big(  \alpha_{\perp}\cdot
	(i \partial_{\perp} + g A_\perp) + \gamma^{0} m_V \big ) \psi_{+}.
\end{equation}

The light-front chiral transformation acts only on $\psi_+$
\begin{equation}
	\psi_+ \longrightarrow \psi_+ + \delta \psi_+ ~~, ~~~ \delta
		\psi_+ = - i \theta \gamma_5 \psi_+,
\end{equation}
and the transformation on $\psi_-$ is given by the equation of
constraint
\begin{eqnarray}
	\delta \psi_- && = { 1 \over i \partial^+}
\big (  \alpha_\perp\cdot(i  \partial_\perp + g A_\perp) + \gamma^{0} m_F \big)
		\delta \psi_{+} \nonumber \\
	&& = - i \theta  \gamma_5 { 1 \over i \partial^+}
\alpha_\perp \cdot \big( i \partial_\perp+ g A_\perp \big) \psi_{+}
		- i \theta m_F \gamma^0 \gamma_5
		{1 \over i \partial^+} \psi_{+}  .
\end{eqnarray}
Thus, the light-front chiral transformation on $\psi$ is
\begin{eqnarray}
	\delta  \psi && = \delta { \psi}_+ + \delta
		{\tilde \psi}_-  \nonumber \\
	&& = -i \theta \gamma_5 \psi_+ - i \theta \gamma_5
		{ 1 \over i \partial^+} \alpha_\perp \cdot
\big ( i \partial_\perp + g A_\perp \big) \psi_{+} - i \theta m_F \gamma^0
		\gamma_5 {1 \over i \partial^+} \psi_{+} . \label{eq.lfct}
\end{eqnarray}

The axial vector current is $j_5^{\mu} =
\bar{\psi} \gamma^{\mu} \gamma_5 \psi$, and the conserved light-front axial
vector charge is
\begin{eqnarray} Q^{5}_{LF}  =  \int dx^- d^2 x_\perp  j^{+}_{5}(x). \end{eqnarray}
Explicit calculation using the field expansions and normal ordering leads to
\begin{eqnarray}
	Q^{5}_{LF} = && \int  {dk^+ d^2 k_\perp \over 2(2 \pi)^3}
		\sum_{\lambda} \lambda \big [ b_{\lambda}^{\dagger}(k)
		b_{\lambda}(k)  + d_{\lambda}^{\dagger}(k)
		d_{\lambda}(k)  \big ] . \nonumber \\
&&  \end{eqnarray}
Thus $Q^{5}_{LF}$ measures the helicity.

What does this teach us? An interaction in which quark helicity changes is
required to remove the $\pi-\rho$ degeneracy produced by massless, canonical
light-front QCD.

There is one explicit chiral symmetry
breaking term in the canonical Hamiltonian,
\begin{eqnarray}
	 g m_V \int dx^- d^2 x_\perp \psi_+^\dagger \sigma_\perp
		 \cdot \left(A_\perp  {1 \over \partial^+}\psi_+
		 - {1 \over \partial^+}(A_\perp \psi_+) \right);
\end{eqnarray}
and this is the only relevant operator that violates light-front chiral
symmetry, although the full interaction contains an unknown function of
longitudinal momenta.

\section{Crude Estimates}

Assume that the $\pi-\rho$ mass splitting is primarily due to
gluon emission/absorption with a helicity flip. This does not
necessarily imply that the pion contains significant glue when the
cutoff is lowered to $\order(1 GeV)$ because renormalization produces a direct
quark-antiquark interaction of the form:
\begin{equation}
V_{\chi SB} = c M^2_{\chi SB} {\bf \sigma_{q \perp} \cdot \sigma_{\qbar
\perp}} \;.
\end{equation}
The real interaction is not proportional to a constant $c$, but this will
give us a crude idea of its effects. The massless spectrum without chiral
symmetry breaking is qualitatively similar to the glueball spectrum. We can
assume an unperturbed mass of $\order(1 GeV)$ for the $\pi$ and the $\rho$.
However, acting on these states, $V_{\chi SB}$ produces:
\begin{equation}
 V_{\chi SB} |\pi> = -{c \over 2} M^2_{\chi SB} |\pi> \;, 
V_{\chi SB} |\rho_0> = {c \over 2} M^2_{\chi SB} |\rho_m> \;,
V_{\chi SB} |\rho_{\pm 1}> = 0 \;;
\end{equation}
where the subscript on the $\rho$ indicates its spin projection.

Since the zeroth-order eigenstates are also eigenstates of this operator, we
can immediately determine the masses,
\begin{equation}
M_\pi^2 = M_0^2 - {c \over 2} M_{\chi SB}^2 \;,
M_{\rho_0}^2 = M_0^2 + {c \over 2} M_{\chi SB}^2 \;,
M_{\rho_{\pm 1}} = M_0^2 \;.
\end{equation}
If we adjust $M_{\chi SB}$ to produce a massless pion, we apparently destroy
rotational symmetry in the $\rho$ multiplet. Violation of few-body
kinematic rotational symmetries is common in light-front field theory, where
rotations about transverse axes involves parton production and annihilation.
If we break rotational invariance in the $q \qbar$ sector without gluons by
100\%, we can restore it when glue is added, but only if the $q \qbar g$
component of the state is comparable to the $q \qbar$ component.

Taking into account that mixing $q \qbar$ and $q \qbar g$ also produces a
negative self-energy, and that the combination must be negative, we find that
this admixture lowers the mass of the $\rho_0$ less than it lowers the masses
of the $\rho_{\pm 1}$ or the $\pi$. The simplest possibility is that the
$\rho$ has a much smaller $q \qbar g$ component than the $\pi$, and that this
component is slightly larger in the $\rho_{\pm 1}$ than in the $\rho_0$. If
this is the case, we can consider this admixture in the $\pi$ alone. 

Assuming
the $q \qbar g$ mass is higher than the $q \qbar$ mass before the chiral
symmetry breaking interaction is turned on, and that confinement produces a
mass gap that is linear in the number of constituents (which may well be
false) and that there is a large gap between the first two $q \qbar g$
states, we can make a crude approximation by considering only one $q \qbar$
mode coupled to one $q \qbar g$ mode. This produces a drastically simplified
hamiltonian (mass-squared operator) of the form,
\begin{equation}
H=\left(\begin{array}{cc}
                M_0^2 & M_0 M_1 \\ M_0 M_1 & M_1^2
                \end{array}\right)~~.
\end{equation}
I have adjusted the off-diagonal matrix element to produce a massless pion,
and we find that
\begin{equation}
P_{q \qbar}={M_1 \over M_0+M_1} \;,
\end{equation}
\begin{equation}
P_{q \qbar g}={M_0 \over M_0+M_1} \;.
\end{equation}
This drastically oversimplified analysis can only lead to $P_{q \qbar} >
P_{q \qbar g}$, which is not necessary, but this type of analysis leads
to the conclusion that the probability of finding glue in the pion may only be
of order $1/2$. Nonetheless, it seems unlikely that any model of the pion that
does not include constituent glue can be derived from QCD, even if it is
possible to derive a valence approximation for other hadrons.

{\bf Acknowledgements:} I am indebted to Roger Kylin for many useful
conversations and for preliminary bound state results with various kinetic and
vertex masses. There are many people who have helped me
understand how the pion might emerge in light-front QCD, including Daniel
Mustaki, Ken Wilson, Avaroth Harindranath, Stanis{\l}aw G{\l}azek, Matthias
Burkardt, Martina Brisudova, and Brent Allen. I am also indebted to Stan
Brodsky for pointing out their work showing that the $q \qbar$ component of
the exact pion is significant. This work was supported by the National Science
Foundation under grant PHY-9800964.



\end{document}